\documentclass[preprint,showpacs]{revtex4}
\usepackage{graphics,graphicx,epsfig}

\begin{document}
\title{Effects of grains' features in surface roughness scaling}
\author{T. J. Oliveira${}^{a)}$ and F. D. A. Aar\~ao Reis${}^{b)}$
\footnote{a) Email address: tiagojo@if.uff.br\\
b) Email address: reis@if.uff.br (corresponding author)}}
\affiliation{Instituto de F\'\i sica, Universidade Federal
Fluminense, Avenida Litor\^anea s/n, 24210-340 Niter\'oi RJ, Brazil}


\begin{abstract}

We study the local and global roughness scaling in growth models with grains at
the film surfaces. The local roughness, measured as a function of
window size $r$, shows a crossover at a characteristic length $r_c$, from a
rapid increase with exponent $\alpha_1$ to a slower increase with exponent
$\alpha_2$. The result $\alpha_1\approx 1$ is explained by
the large height differences in the borders of the grains when compared to
intragrain roughness, and must not be
interpreted as a consequence of a diffusion dominated intragrain dynamics.
This exponent shows a weak dependence on the shape and size distribution of the
grains, and typically ranges from $0.85$ for rounded grain surfaces to $1$ for
the sharpest ones. The scaling corrections of exactly solvable models
suggest the possibility of slightly smaller values due to other smoothing
effects of the surface images. The crossover length $r_c$
provides a reasonable estimate of the average grain size in all model systems,
including the cases of wide grain size distributions. In Kardar-Parisi-Zhang
growth, very different values
of $\alpha_2$ are obtained, ranging from $0.4$ for the films with smoothest
surfaces to values in the range $0.1\lesssim \alpha_2 \lesssim 0.2$ for systems
with large cliffs separating the grains. Possible applications to real systems
which show this crossover with similar exponents are discussed.

\end{abstract}

\maketitle

\section{Introduction}
\label{intro}

In the last twenty years, many works on thin film and multilayer growth analyzed
the scaling properties of the local surface roughness (interface
width) $w$, which represents height fluctuations in different lengthscales, and
of the global roughness $\xi$, which represents the height fluctuations of the
full system. Alternatively, some works considered height-height correlation
functions, which exhibit the same scaling of the local roughness. In a certain
growth process, the measured set of scaling exponents
suggests its association with some universality class, consequently it is
a useful tool to find the dominant growth mechanisms
\cite{barabasi,krug,krim,evansreview}.
However, simple scaling features \cite{famvic} are observed only in a small
number of real systems, as well as in a small number of growth models
\cite{chamereis}. Instead, a crossover in the roughness scaling is frequently
observed as the window size $r$ or the time $t$ varies
\cite{lita,vasco,kleinke,ebothe,vazquez,marta,tersio,mendez,otsuka,hiane,nara,auger,pehlivan}.
This crossover may be related to the presence of grains at the film surface
\cite{lita,vasco,kleinke,ebothe,vazquez}, with the typical size dependence of
the local roughness $w$ illustrated in Fig. 1
($w$ is an average over windows of size $r$ gliding through a much larger
surface). For small $r$, one obtains $w\sim r^{\alpha_1}$ with
$\alpha_1\sim 1$. For large $r$, $w\sim r^{\alpha_2}$ is obtained, with
$\alpha_2$ typically ranging from $0$ to $0.5$ (in electrodeposition works,
$\alpha_2>0.5$ is also frequent). When $r$ is of the order of the lateral
correlation length, $w$ saturates at $w=\xi (t)$. The expected time
evolution $\xi\sim t^{\beta}$ may also show a crossover, i.e. different
values of $\beta$ may be found in different time regions.

It is usually suggested that the characteristic length $r_c$ of the first
crossover in Fig. 1 is of the same order of the average grain
size. Moreover, the usual interpretation for $\alpha_1\approx 1$ is that the
small lengthscale features are related to a diffusion dominated, Wolf-Villain
(WV) type growth \cite{barabasi,wv}. At longer lengthscales, the height
fluctuations follow a different dynamics. When $\alpha_2\approx 0.4$ is
obtained, it is usually associated with the Kardar-Parisi-Zhang (KPZ) growth
\cite{kpz} - see also Ref. \protect\cite{kpz2d}. The
interpretation of smaller but nonzero values of $\alpha_2$ is not so clear.
However, when $\alpha_2\approx 0$, the process is associated with
the Edwards-Wilkinson (EW) class \cite{ew}.

Despite the large number of works showing the roughness scaling of Fig. 1, it
seems that no systematic
analysis of these apparently universal features in surface growth models has
been presented yet. As far as we know, this crossover was illustrated only with
a very simple model of grain aggregation in Ref. \protect\cite{vazquez}, while
a crossover in the exponent $\beta$ was discussed in Ref.
\protect\cite{albano}.
The aim of the present work is to fill this gap by studying statistical growth
models with grains at the film surface, focusing on the effects of the
shapes of the grains, their average size and their size distributions. Our first
step is to show
that the result $\alpha_1\approx 1$ is a consequence of the large height
differences between neighboring grains when compared to the intragrain height
fluctuations. Thus, this result is not related to the intragrain growth
dynamics. Subsequently, the effects of the shape of the grains,
of their average size and size distribution are addressed by more refined
models of grain aggregation (the simplest ones inspired in the model of Ref.
\protect\cite{vazquez}). The qualitative behavior illustrated in Fig. 1 is
obtained in all models, with $\alpha_1\lesssim 1$, and $r_c$ is shown to be an
accurate estimate of the average grain size. The effect of size distribution is
relatively weak, but rounded grain surfaces may reduce $\alpha_1$ to values
near $0.85$ and rounded grain boundaries may play the same role. We also discuss
other nontrivial features, such as the small effective exponents $\alpha_2$ in
some KPZ models, which may be relevant for some applications.

In all models presented here, the microscopic mechanisms of grain
formation are not represented, but only the intergrain aggregation properties.
We are implicitly assuming that the characteristic time in which a
grain is formed is much smaller than the time interval between aggregation of
two neighboring grains. Other models of polycrystalline growth
\cite{albano,huang,rubio} also adopt different
stochastic rules to form the grains and to generate large scale correlations.
Despite this limitation, such an approach is essential to analyze large
three-dimensional deposits with reasonable computational effort. Also notice
that our growth models are analyzed in the growth regime ($t/L^z\ll 1$,
where $L$ is the large substrate length and $z$ is the growth exponent), which
parallels the typical experimental conditions.

The rest of this work is organized as follows. In Sec. II we analyze simple,
exactly solvable models with grains at the surface, which explains the result
$\alpha_1\approx 1$. In Sec. III we analyze the effects
of the shape of the grains using models with grains of fixed size and
constraints in the local height differences. In
Sec. IV we present the models for deposition of grains with different size
distributions and give evidence of their KPZ scaling in the continuous limit.
The scaling of the local and global roughness of these models are studied in
Sec. V, where we also discuss possible applications. In Sec. VI we summarize our
results and present our conclusions.

\section{The origin of the crossover with $\alpha_1 =1$}
\label{checkerboard}

Here we consider simple but instructive models with grains regularly placed in
infinitely large lattices, with the unit lattice parameter corresponding to the
size of an image pixel. They have the advantage of being exactly solvable and
of easy interpretation (a related model is used in Ref.
\protect\cite{chatraphorn} for calculation of height-height correlations). For
three-dimensional
structures, the local roughness $w(r)\equiv {\langle
{\overline{{\left( h-\overline{h}\right)}^2}}^{1/2}\rangle}$ is the
configurational average (angular brackets) of the root mean squared (rms) height fluctuation
in square windows (overbars) of integer size $r$ which span the whole surface of
the deposit. For two-dimensional structures, the
one-dimensional windows are linear segments of size $r$.

The first model is the square wave, shown in Fig. 2a, where the grains are
rectangles of width $l$ and height $H$. This structure
contains the most important geometric feature of a deposit with grains at the
surface: The height difference between bordering sites is much larger than the
roughness of the surface of a grain (top of the grain).
Due to the translation invariance, it is relevant to
calculate the roughness only for $r<2l$.
In the interesting limit $r\gg 1$, the sums over the possible window positions
are turned into integrals and lead to
\begin{equation}
w_{sw} = \frac{\pi}{8} H \frac{r}{l} .
\label{wsquarewave}
\end{equation}
When the window size varies and $l$ and $H$ are kept fixed, we obtain $w\sim r$,
consequently $\alpha =1$. This linear dependence follows from the contribution
of the windows that partly cover one grain (height
$H$) and partly cover the substrate (zero height), since the fraction of
windows with this property is $r/l$, i.e. it is proportional to $r$.

In order to show that this result is also valid in three-dimensional
systems, we also calculated the local roughness in the structure shown in
Fig. 2c: The grains are blocks of lateral size $l$ and height $H$, which are
formed by cubic units of size $1$ (the unit lattice parameter). We obtain 
\begin{equation}
w(r) =  \frac{\pi}{4} \frac{r}{l}  H {\left[ 1-
\left( \frac{3}{4}-\frac{\ln2}{2}\right) \frac{r}{l}\right]} .
\label{wchecker}
\end{equation}
It also gives a dominant contribution as $w\sim Hr/l$, thus $\alpha =1$.
Again, this contribution comes from the windows that partly cover one grain
(height $H$) and partly cover the substrate (zero height), whose number is
proportional to $r$.

Equation (\ref{wchecker}) shows that the deviation from the
linear relation between $w$ and $r$ is determined by a factor near
$0.40\frac{r}{l}$. Thus, even for $r/l\approx 1/5$, the deviations are below
$10\%$. Indeed, only small deviations from $\alpha_1=1$ are found in the
three-dimensional growth models presented in Secs. III-V.

However, there is a weakness in the above model systems: the images
obtained from microscopy techniques usually do not show the steep surface
features of Fig. 2a, even if they are present in the
real surfaces. Instead, the surfaces usually look smoother in the final images.
For this reason, we also analyzed the structure shown in Fig. 2b: within a
region of size $a$, with $a\ll l$, the local height varies from the lowest to
the highest value. This "smoothened square wave" was also discretized in order
to adopt the same scheme for the calculation of the roughness. Assuming $a\gg 1$
and $r\gg1$ for converting the sums into integrals and expanding in powers of
$a/l$, we obtain after some algebra
\begin{equation}
w_{ssw} = \frac{\pi}{8} H \frac{r}{\left( l+a\right)} \left[ 1 +
\left( \frac{32\sqrt{3}}{15\pi} - \frac{8\sqrt{2}}{3\pi} \right)
{\left( \frac{a}{r}\right)}^{3/2} + o{\left( \frac{a}{r}\right)}^{5/2} \right] .
\label{wsmoothwave}
\end{equation}
It also shows a dominant term with $w\sim r$, i.e. with $\alpha_1 =1$. The
correction term is proportional to ${\left( \frac{a}{r}\right)}^{3/2}$, with
negative coefficient of order $-0.1$. This correction is small even when $r\sim
a$, but the negative value suggests that the effective exponent $\alpha_1$ may
be slightly below one under these conditions. In real images, this result may be
important for the smallest window sizes.

The above analysis shows that $\alpha_1=1$ is a consequence of two simple facts:
The contribution of the grain borders dominates the average roughness ($w\sim
H$ at those regions) and the fraction of windows with this property is
proportional to $r$. Consequently, this roughness exponent must not be
associated with a particular intragrain growth dynamics.

At this point, it is important to recall that a diffusion dominated growth with
negligible nonlinear effects also has roughness exponent $\alpha=1$ in $2+1$
dimensions. This system is described by the linear molecular-beam epitaxy (MBE)
equation (or Mullins-Herring equation) \cite{mullins,herring}
${{\partial h}\over{\partial t}} = \nu_4{\nabla}^4 h + \eta (\vec{x},t)$
($\nu_4$ is a constant) in the hydrodynamic limit. Certainly our results do not
discard this possibility in a real system, but our conclusion is that simple
dimension-independent geometric features are sufficient to explain
$\alpha_1\approx 1$ for a system
with grains at the surface, regardless of the intragrain dynamics.

A possible suggestion to identify the origin of this scaling exponent is to
analyze the average squared roughness $w_2 \equiv \left< { \left( h -
\overline{h}\right) }^2 \right>$. It differs from $w$ only by the stage of the
averaging process in
which the square root of a height fluctuation is calculated, but both are
expected to have the same scaling properties at long
lengthscales. For the system of Fig. 2c, we obtain
\begin{equation}
\sqrt{w_2\left( r\right) } = \frac{1}{\sqrt{6}} H {\left( \frac{r}
{l+a}\right)}^{1/2} {\left( 1-\frac{a}{r} + \dots \right)} .
\label{w2checker}
\end{equation}
A log-log plot of the rms roughness $\sqrt{w_2}$ versus $r$ gives $\alpha_1=1/2$
because the factor $r/(l+a)$ also appears in the expression of $w_2$ as a
consequence of the configurational average. The main problem of studying this
quantity is that $\alpha_2$ is close to $1/2$ in several systems, such as KPZ
ones, thus other corrections to scaling may hide the crossover.

\section{Effects of the shape of the grains on the local roughness}
\label{shape}

Now we consider the first set of three-dimensional models for thin film growth
with the presence of grains at the surface. This type of model was formerly
proposed in Ref. \protect\cite{vazquez}. The first step is to
grow a deposit with cubic particles of unit size following the rules of some
simple growth model. The first one is the restricted solid-on-solid (RSOS)
model, in which the aggregation of the incident
particle is accepted only if the heights differences of nearest neighbors are
always $0$ or $1$, otherwise the aggregation attempt is rejected \cite{kk}. The
second one is the Family model, in which the incident particle aggregates at
the column with the smallest height among the incident column and its nearest
neighbors \cite{family}. In both cases, deposits with small local slopes are
formed. After growing the film with one of these rules, the size of
each particle is enlarged by a factor $l$, i.e. each particle is transformed
in a cubic grain of size $l$. Finally, the upper parts of the grains at the
surface of the film are polished with the shape of a sphere of radius $R$, as
shown in Fig. 3a. The cross section of the resulting grains for $l=32$ and
several values of $R$ are shown in Fig. 3b. A cross-sectional view of the
surface of a deposit grown with the RSOS model and $l=32$ is shown in Fig. 3c.

Square substrates of lateral length $128$ are
considered for both models, with deposition times below $400$ (in units of
number of deposition attempts per substrate site), which ensures that the
systems are in the growth regime ($t/L^z\ll 1$). The lateral length of the
final lattice with grains is $128 l$.

In the hydrodynamic limit, the RSOS model is
described by the KPZ equation, in which the local height $h(\vec{x},t)$ evolves
as ${{\partial h}\over{\partial t}} = \nu{\nabla}^2 h + {\lambda\over 2}
{\left( \nabla h\right) }^2 + \eta (\vec{x},t)$
($\nu$ and $\lambda$ are constants and $\eta$ is a Gaussian noise). The Family
model is described in that limit by the EW equation, which corresponds to the
KPZ equation with $\lambda =0$.

In Fig. 4 we plot the local roughness as a function of window radius for the
RSOS model with $l=32$ and two values of $R$, as well as the case with
nonpolished,
cubic grains ($R\to\infty$). Even with smooth grain surfaces and the constraint
in the local slopes between the grains, the crossover in the local roughness
scaling is present. Estimates of the
crossover lengths $r_c$ are obtained from the intersections of linear fits in
two scaling regions (see illustration of the method in Fig. 1). In all cases
shown in Fig. 4,
$r_c$ is very close to the grain size $l$; when $8\leq l\leq 32$ is considered,
the maximum difference between $l$ and $r_c$ is $5\%$, and this difference tends
to decrease for larger $l$.
The average slopes of the curves before the crossover ($\alpha_1$) range from
$1$ for the cubic grains to $0.85$ for the smoother grain surfaces
($R=22$). After the crossover, we obtain $\alpha_2\approx 0.4$ for all shapes,
which is in good agreement with the best currently known
estimates $\alpha\approx 0.39$ for the KPZ class \cite{kpz2d}.
The same qualitative results were obtained with the Family model, but with
$\alpha_2\approx 0$ after the crossover, which is hard to
distinguish from the saturation at the global roughness $\xi$.

Some experimental works presented values of $\alpha_1$ and $\alpha_2$ close to
the above ones for the KPZ model \cite{lita,kleinke,marta,nara}, while slightly
larger values of $\alpha_2$ were obtained in electrodeposition works of Refs.
\protect\cite{vazquez,otsuka,hiane}. Our results
confirm the interpretation of asymptotic KPZ scaling in the cases
where $\alpha_2\approx 0.4$ was obtained. The formation of grains or other
surface structures separated by cliffs are responsible for the crossover.
However, one important conclusion of the above analysis is that their average
sizes may be estimated with reasonable accuracy from the crossover lengths
$r_c$ calculated with the procedure shown in Fig. 1, independently of the shape
of the grains. This is reinforced by results for other models in Sec. V.
Finally, the above results show that the conclusion of Sec. II can be extended
to surfaces with large lengthscale fluctuations: $\alpha_1\approx 1$ is related
to simple geometric features and not to the intragrain dynamics. Thus, in a
particular application, this dynamics remains to be clarified by specific
models accounting for the interactions between individual atoms or molecules.

\section{Models of grain aggregation and their universality class}
\label{models}

Now we introduce a more realistic model for grain aggregation, which, in
contrast to the model of Sec. III, represents deposition of
grains of different sizes and disorder in their aggregation positions. This
section is devoted to the presentation of the model and the study of its
universality class. The discussion of local and global roughness scaling, as
well as possible applications, is done in Sec. V.

The model is defined in a simple cubic lattice where the length unit is the
lattice parameter. The grains have cubic shapes and the lateral size $l$ of
each one is chosen from a certain distribution. They sequentially incide
perpendicularly to an initially flat substrate, with two of their faces parallel
to the substrate. The incident grain permanently
aggregates to the deposit when its bottom touches a previously aggregated grain
(thus, there is no lateral aggregation). The process is illustrated in Fig. 5a.
Three grain size distributions are considered here: Delta distribution (fixed
grain size), Poisson distribution with average size $\overline{l}$, and
Gaussian distribution with average size $\overline{l}$ and width
$\overline{l}/3$. The Poisson and Gaussian distributions with $\overline{l}=32$
are illustrated in Fig. 5b. In most simulations in the growth regime (small
depositon times), substrate sizes $L=4096$ will be considered. The time unit is
chosen to correspond to the deposition time of one monolayer of individual
atoms, i.e. $L^2$ atoms.

Similarly to the previous models, this one does not account for the details of
the grain formation, but assumes that it takes place in a relatively small time
interval, in which the atoms arriving in a region of size $\sim l$
diffuse in this neighborhood until finding a final position of aggregation. On
the other hand, this model leads to the formation of a porous deposit which
resembles those produced with Tetris model for compaction of granular media
\cite{tetris} or ballistic deposits \cite{vold}. However,
we are only interested in surface features and, consequently, the main aspect to
be considered here is the generation of intergrain correlations along the
surface. For instance, notice that the aggregation of a new grain only depends
on the current surface features and not on the bulk properties.
The size distributions considered here are also not expected to be
representative of real growth processes (see e. g. Refs.
\protect\cite{weaire,castro}), but just designed to the investigation of
qualitative effects.

Due to the formation of pores in the deposit, it is natural to expect that this
model is in the KPZ class for any grain size distributions. This is confirmed by
the comparison of global roughness distributions at the steady states ($t/L^z
\gg 1$) with other KPZ models - see e. g. an application of the method for
ballistic deposition in Ref. \protect\cite{distrib}. In Fig. 6 we show the
steady state squared roughness distributions of the Poisson and the Gaussian
models with $\overline{l}=8$ and $L=1024$ and the distribution for the original
RSOS model with $L=256$, which is 
the most accurate representative of the KPZ class in $2+1$ dimensions
\cite{distrib}. There, $P\left( \xi^2\right) d\xi^2$ is the probability that the
global square roughness is in the range $\left[ \xi^2, \xi^2+d\xi^2\right]$,
$\langle \xi^2\rangle$ is the average squared roughness and $\sigma$ is the rms
deviation in each distribution. The good data collapse of the normalized
distributions suggests that the grain models are also in the KPZ class. The
estimates of their skewness and kurtosis provide additional support:
$S_{Poisson}=1.66\pm 0.07$, $S_{Gaussian}=1.65\pm 0.08$, $Q_{Poisson}=5.2\pm
0.7$ and $Q_{Gaussian}=5.1\pm 0.6$, which are very close to the values
$S=1.70\pm 0.02$ and $Q=5.4\pm 0.3$ obtained with the RSOS and other KPZ models
\cite{distrib}. Results for smaller $\overline{l}$ confirmed this result with
higher accuracy. Results for larger $\overline{l}$ are not so accurate because
it is difficult to attain steady states in lattices that are large enough to
satisfy the condition $\overline{l}\ll L$ (this is a necessary condition when
we search for a continuous description of the discrete model).

\section{Effects of grain size and size distribution on local and global
roughness}
\label{sizeeffects}

Here we analyze the global and local roughness scaling in the models presented
in Sec. IV, focusing on the features of the growth regime, where correlation
lengths along the surface are much smaller than the lateral system sizes.

In Fig. 7 we show the time evolution of the global roughness $\xi$ for the
models with Poisson, Gaussian and delta distributions with average grain size
$\overline{l}=32$. There is no qualitative difference in these curves, but only
a small shift to larger values of $\xi$ due to the presence of larger grains in
the Gaussian case, followed by the Poissonian case (see Fig. 5b). Up to
$t\approx 150$, there is a rapid increase of the roughness, approximately as
$\xi\sim t^{0.9}$. This roughly corresponds to five layers of grains. A
growth exponent $\beta\approx 1$ at small times is typical of a columnar
growth, where some points of the substrate are exposed while the average height
is linearly increasing in time.
For longer times, $\xi$ is slowly increasing with exponent $\beta\approx 0.1$,
which is much lower than the KPZ value $\beta\approx 0.23$ \cite{kpz2d}. No
crossover to the asymptotic KPZ value is observed in our simulations
up to $t\sim {10}^4$, even for smaller values of $\overline{l}$.

In Fig. 8 we show the local roughness $w$ as a function of the window size $r$
at $t=2000$ for the same distributions of grain sizes. We do not identify long
regions with a power-law growth of $w(r)$ in that plot, but only approximate
power laws for regions of small and large $r$, as illustrated by the dashed
lines of Fig. 8. For small $r$, we obtain $\alpha_1\approx 1$, independently of
the average grain size and grain distribution, as expected from the sharp
shape of the grains (see Secs. II and III). All estimates of $r_c$ are
very close to $\overline{l}$; for instance, in Fig. 8 we have $\overline{l}=32$
and $r_c=30\pm 2$. This is confirmed by results for other grain size
distributions. Moreover, slightly different
choices of the regions for the linear fits (dashed lines in Figs. 1 or 8) do not
give very different values of $r_c$. These results confirm that the crossover
length $r_c$ (obtained with the procedure illustrated in Fig. 1) is a
reliable estimate of the average grain size, and not only an estimate of its
order of magnitude.

However, we obtain $\alpha_2\approx 0.1$ in Fig. 8, which is much lower than the
KPZ exponent $\alpha\approx 0.39$.
Simulation for longer times and small $\overline{l}$ shows that a slow increase
of the slope of the $\log{w}\times\log{r}$ plot takes place. This is
illustrated in Fig. 9, where we plot the local roughness for $\overline{l}=8$
(Poisson distribution) at time $t=6000$: after the crossover at $r_c$, the local
slope of the plot increases from $0.1$ to $0.18$ within two decades of $r$.
Thus we conclude that the exponents calculated from roughness scaling in these
systems are not consistent with their true asymptotic class, which was shown to
be KPZ in Sec. IV. Instead, the small values of $\alpha_2$ may lead to a
confusion with the logarithmic scaling ($\alpha=0$) of linear EW growth, which
would be an absolutely wrong interpretation of the growth dynamics.

The inefficiency of local roughness scaling to show the true universality class
of a
growing system was already observed in several situations \cite{chamereis}. In
the present case, the explanation for the discrepancies seem to be the presence
of large local slopes at the borders of the grains. In Figs. 10a and 10b, we
show cross-sectional views of the surfaces of deposits
grown with the grain deposition model and with the ballistic deposition model
\cite{barabasi,vold}, which has similar surface features. They look much rougher
than the surface of the KPZ model of Fig. 3c (based on RSOS growth)
because the height differences were limited to the grain size in that system. On
the other hand, we recall that the finite-size estimates of roughness exponents
of ballistic deposits are also much smaller than the KPZ values
\cite{eytan,haselwandter,balfdaar}. This comparison reinforces the
interpretation that the deviations of $\alpha_2$ from the KPZ value in the
grain deposition models are related to the large surface cliffs, whose depths
are frequently much larger than the grain sizes (Fig. 10a).

Instead of being viewed as a limitation of our model, these deviations establish
an interesting point for investigation because similar estimates of $\alpha_1$
and $\alpha_2$ are obtained in some real processes. Examples are
films of $ZnO$ deposited by spray pyrolysis \cite{ebothe}. For low flow rates,
the scaling of the local roughness gives $\alpha\approx 0.4$,
which suggests KPZ scaling. However, for high rates, two scaling
regions of the local roughness are observed, with $0.91\leq\alpha_1\leq 0.95$
and $0.15\leq\alpha_2\leq 0.16$. We believe that this result does not
discard the possibility of KPZ scaling for high flow rates. Instead, the
deviation from the KPZ value may be a consequence of the presence of grains and
an intergrain KPZ dynamics leading to a surface geometry similar to the present
model (although the details of the dynamics are probably very different there).

Our results also suggest a careful interpretation of estimates $\alpha_1\approx
1$ and $\alpha_2\approx 0$. The latter is usually interpreted as a signature of
EW scaling (logarithmic growth). However, it is
reasonable to expect that small data fluctuations may give a power law with a
small $\alpha_2$ instead of a logarithmic growth, particularly when small
scaling regions are considered. Some systems which present these exponents
estimates are pulsed laser deposition of perovskites, in which they were
obtained from power spectral density (PSD) scaling \cite{vasco}, and cooper
electrodeposition with additives \cite{mendez}. In the latter, $\alpha_1\approx
0.86$ and $\alpha_2\approx 0$ were obtained in two scaling regions which are
separated by an intermediate regime with intermediate slope in the
$\log{w}\times\log{r}$ plot. We believe that the investigation of a possible
KPZ scaling to represent the long range correlations is justified in these and
related systems.

\section{Conclusion}

We studied models of thin film growth with a focus on the effects of grain
shape, average size and size distribution on the scaling of local and
global roughness. They were simplified in order to avoid the complexities of the
dynamics of grain formation and to allow the simulation of three-dimensional
deposits with large system sizes.
The local roughness $w$ shows a crossover from a rapid increase with the window
size $r$, for small $r$, to a slower increase for larger $r$, respectively with
effective roughness exponents $\alpha_1$ and $\alpha_2$. The result
$\alpha_1\approx 1$ is explained by
the large height differences in the borders of the grains and must not be
interpreted as a consequence of a diffusion dominated intragrain dynamics,
although the latter gives the same roughness exponent. The value of $\alpha_1$
shows a weak dependence on the shape and size distribution of the grains,
typically varying from $0.85$ for the smoothest grain surfaces to $1$ for the
sharpest ones. The scaling corrections of exactly solvable models suggest that
slightly smaller values may be obtained with pronounced rounding of the grain
boundaries. The crossover length $r_c$ provides reasonable estimates of the
average grain size even in the case of large size distributions.
On the other hand, large ranges of values of $\alpha_2$ are
obtained in KPZ systems: when the height difference between the grains is
limited, $\alpha_2$ is near the KPZ value $0.39$, but systems with large local
slopes (typically with large cliffs separating the grains) show $\alpha_2$ in
the range $[0.1,0.18]$.

The observed crossover from $\alpha_1\approx 1$ to $\alpha_2\approx 0.4$ is
observed in several real growth processes, which is an evidence of KPZ scaling.
However, much smaller values of $\alpha_2$ are also found in some systems. Our
results suggest that this may be an effect
of the presence of grains with large height differences between them, while the
long range correlations are still governed by the KPZ equation. Thus, we
believe that this work motivates more realistic approaches to model growth
processes with formation of grains. For each particular process, it may provide
a better knowledge of the crossover in roughness scaling and the exponents
involved there. They also motivate the analysis of other quantities for a more
reliable test of the universality class of growing systems, such as height or
roughness distributions \cite{distrib}.

Finally, it is important to recall that the models presented here are not cases
of anomalous \cite{lopez,schwarzacher} nor super rough scaling \cite{auger}, in
which the global roughness exponents are very large ($\alpha_g\geq 1$). Indeed,
KPZ or EW systems do not show anomalous scaling \cite{lopez}. Again, we
stress the conclusion that the explanation to the crossover observed in deposits
with grain's features is much simpler, since it is related to the
configurational average in the calculation of the local roughness.

\acknowledgments

TJO acknowledges support from CAPES and FDAAR acknowledges support from CNPq and
FAPERJ (Brazilian agencies). FDAAR thanks Prof. M. U. Kleinke and Prof. R.
Priolli for helpful discussion.

\newpage


\newpage

\begin{figure}[!h]
\begin{center}
\includegraphics[width=13cm]{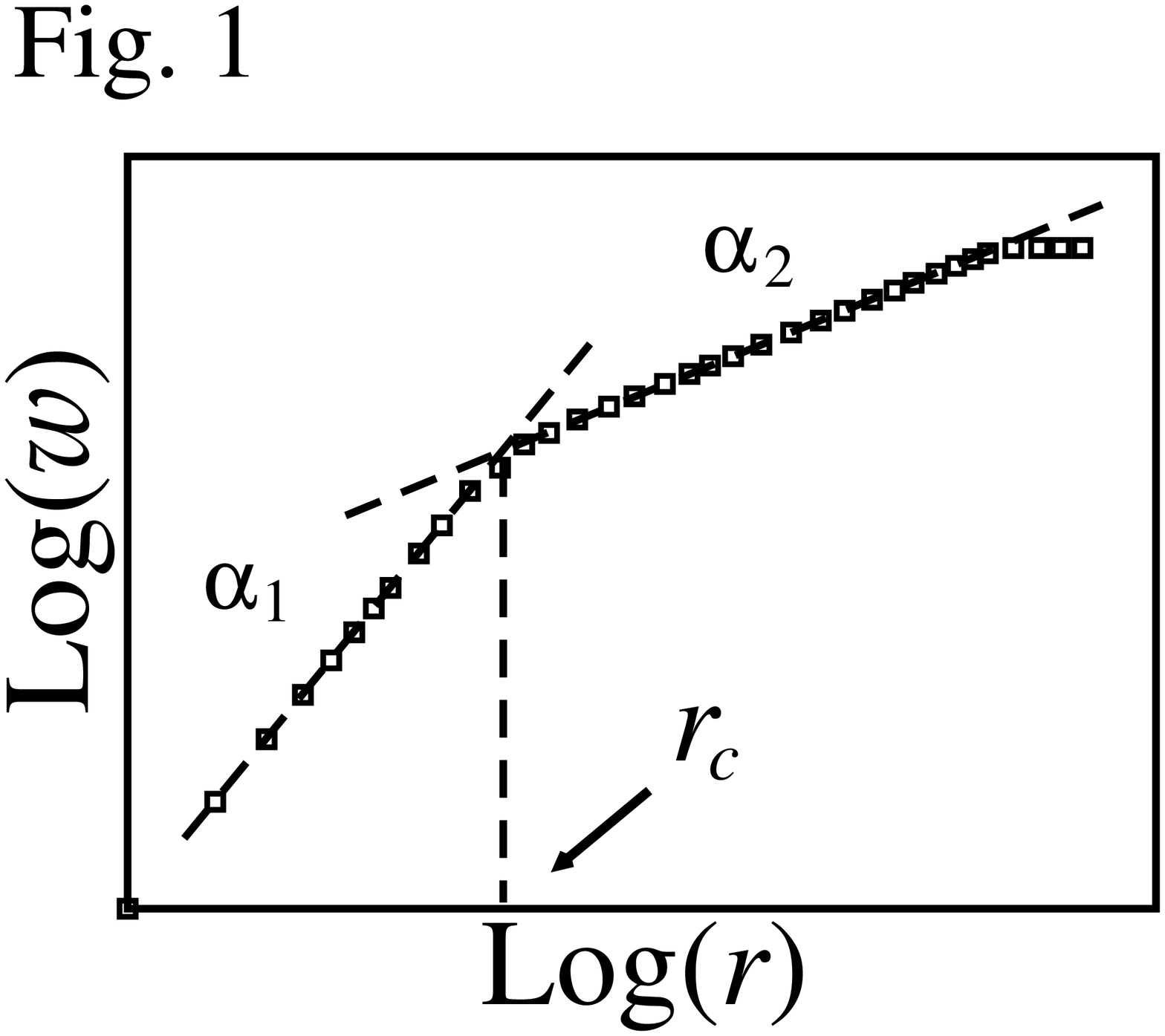}
\caption{Typical dependence of the local roughness on the window size, with
linear fits in two scaling regions (defining exponents $\alpha_1$ and
$\alpha_2$) and a crossover length $r_c$ given by the intersection of those
fits.}
\label{fig1}
\end {center}
\end{figure}

\begin{figure}[!h]
\begin{center}
\includegraphics[width=13cm]{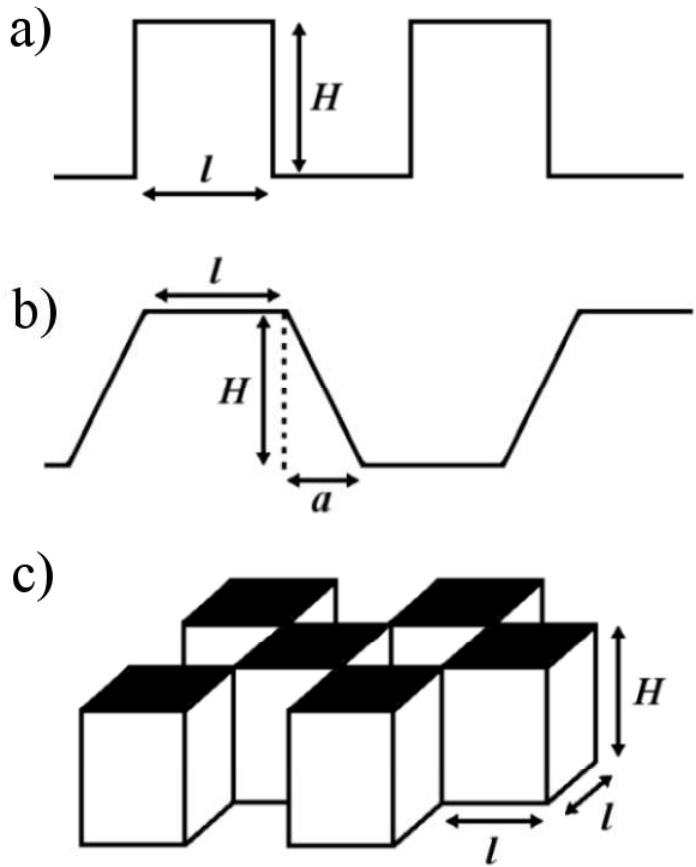}
\caption{Periodic structures with grains of simple shapes: (a) Square wave; (b)
smoothed square wave (trapezoidal wave); and (c) a chekerboard distribution of
blocks.}
\label{fig2}
\end{center}
\end{figure}

\begin{figure}[!h]
\begin{center}
\includegraphics[width=13cm]{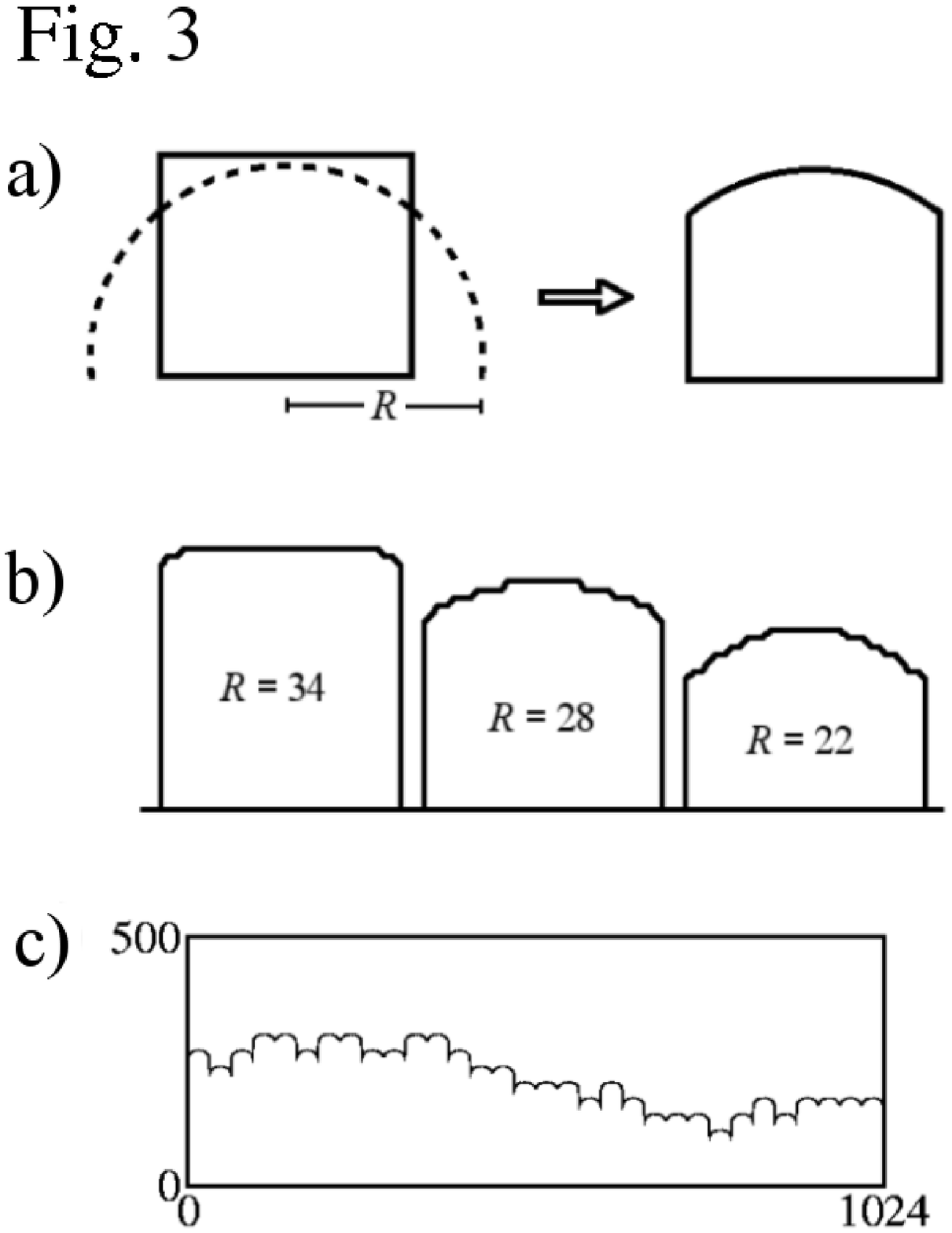}
\caption{(a) Illustration of the method to polish the surface of a cubic grain.
(b) Grains of size $l=32$ with polished surfaces of different radii $R$. (c)
Cross-sectional view of a deposit generated with the RSOS model after
transformation of the unit size surface particles into rounded grains ($l=32$,
$R=22$).}
\label{fig3}
\end{center}
\end{figure}

\begin{figure}[!h]
\begin{center}
\includegraphics[width=13cm]{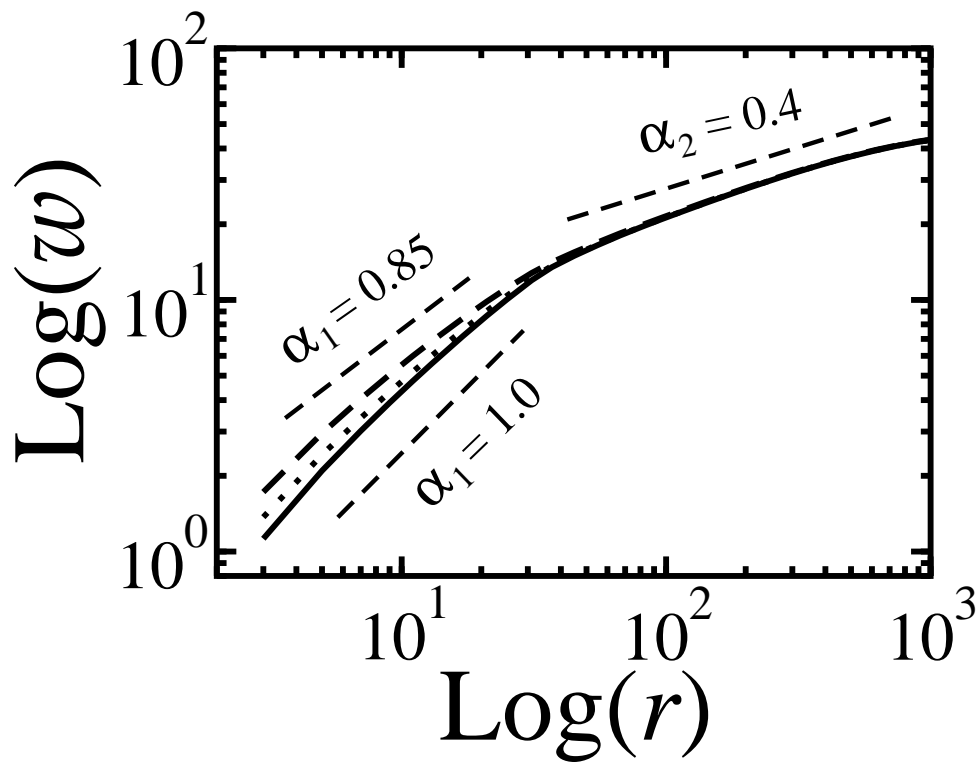}
\caption{Local roughness scaling for the RSOS model with square grains of size 
$l=32$ (solid curve) and rounded grains of the same size and $R=34$ (dotted
line) and $R=22$ (dashed line). Tilted dashed lines show the exponents obtained
in linear fits of each scaling region.}
\label{fig4}
\end{center}
\end{figure}

\begin{figure}[!h]
\begin{center}
\includegraphics[width=13cm]{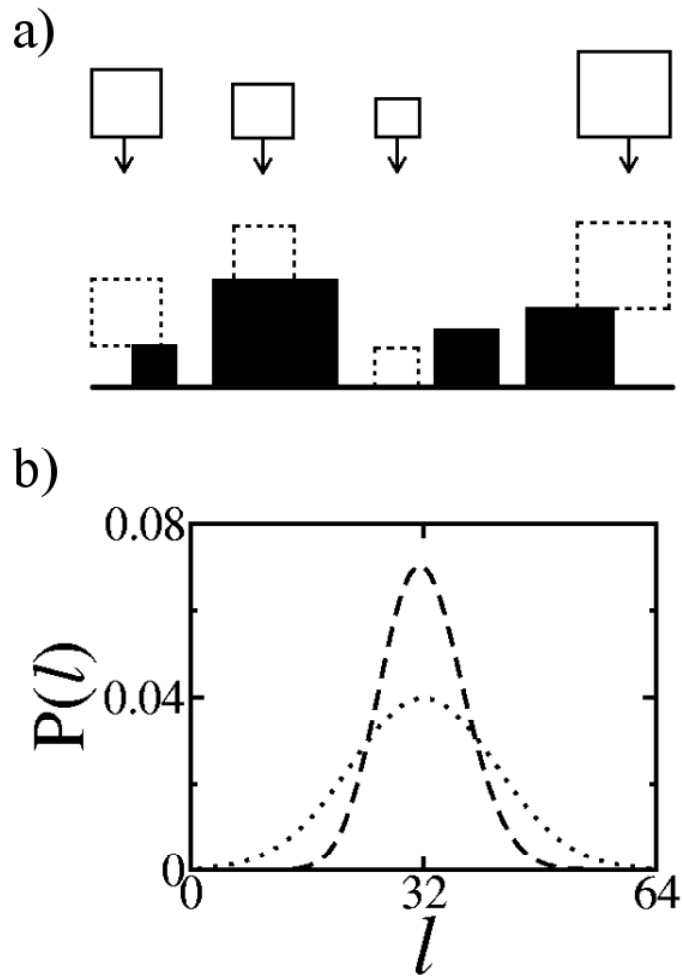}
\caption{(a) Illustration of the deposition rules of the grain aggregation
model. Shaded squares are previously deposited grains, open squares are
incident grains and open dashed squares show their final aggregation positions
(b) Poisson (dashed curve) and Gaussian (dotted curve) grain size distributions
for $\langle l\rangle = 32$.}
\label{fig5}
\end{center}
\end{figure}

\begin{figure}[!h]
\begin{center}
\includegraphics[width=13cm]{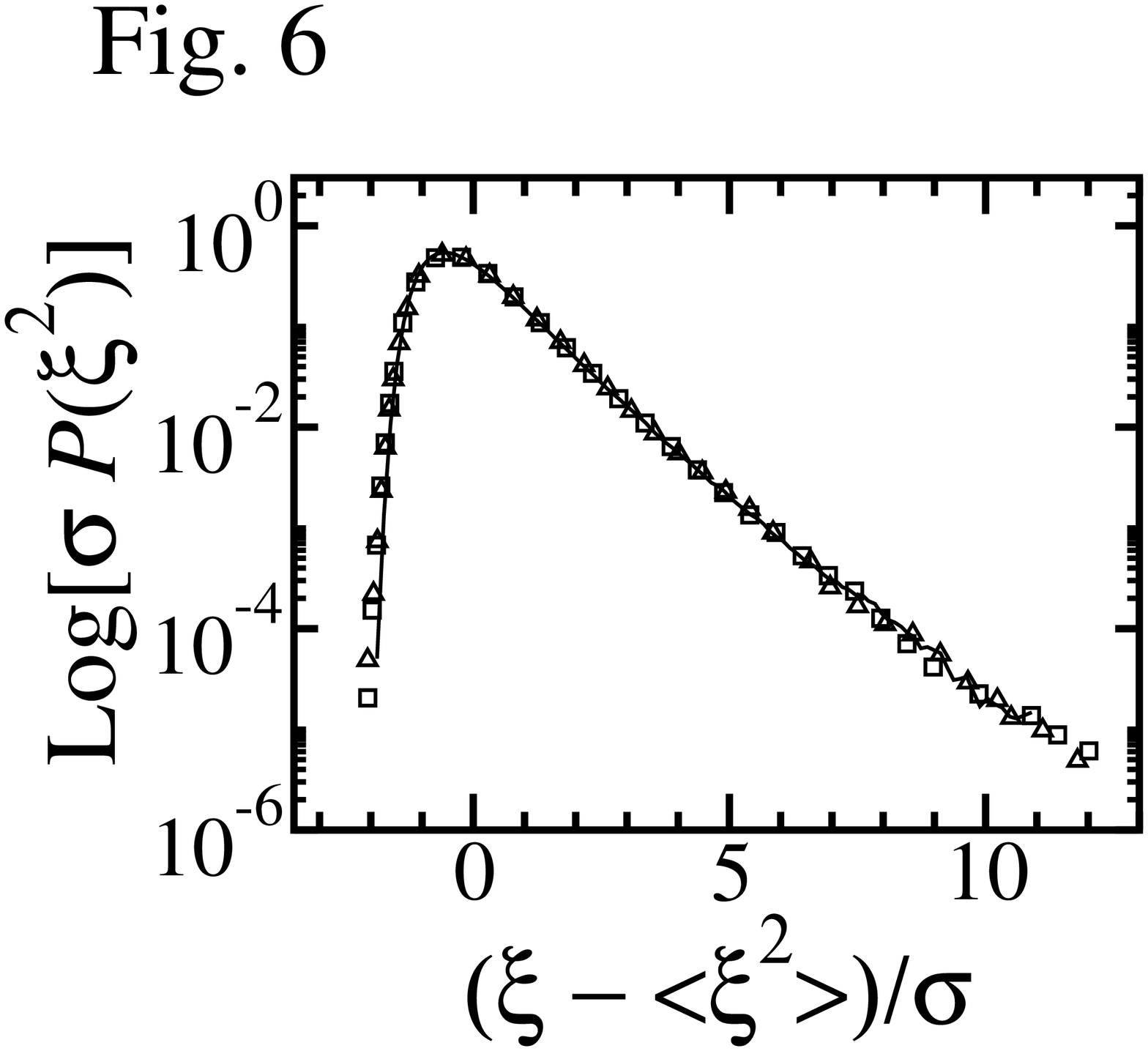}
\caption{Steady state normalized square roughness distributions of the grain
deposition model with $\langle l\rangle =8$ and Poisson (triangles) and Gaussian
(squares) grain size distributions, and of the RSOS model (solid curve).}
\label{fig6}
\end{center}
\end{figure}

\begin{figure}[!h]
\begin{center}
\includegraphics[width=13cm]{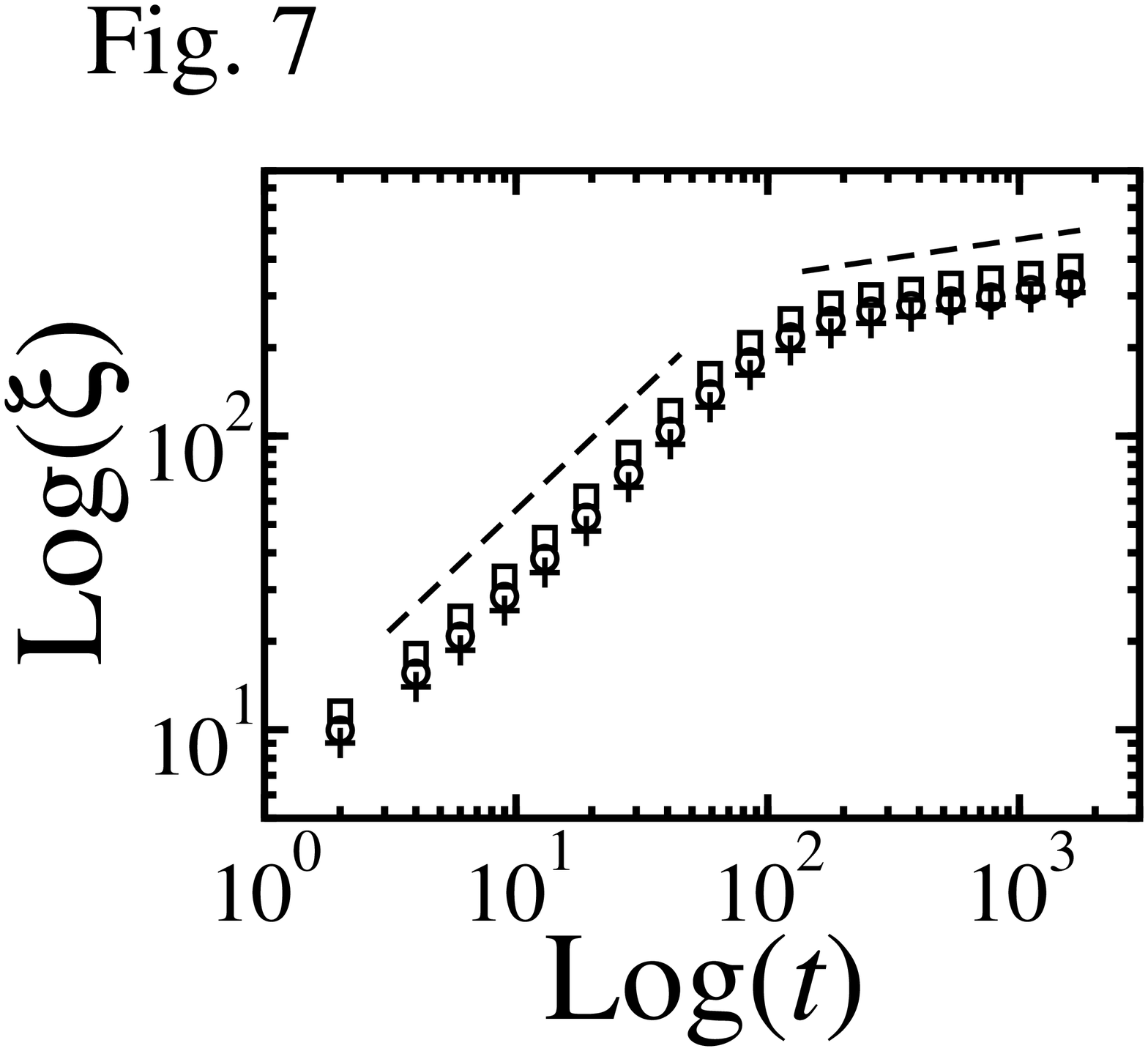}
\caption{Time evolution of the global roughness of the grain deposition model
with $\langle l\rangle =32$: delta (crosses), Poisson (circles) and Gaussian
(squares) size distributions. The dashed lines are parallel to linear fits of
two scaling regions.}
\label{fig7}
\end{center}
\end{figure}

\begin{figure}[!h]
\begin{center}
\includegraphics[width=13cm]{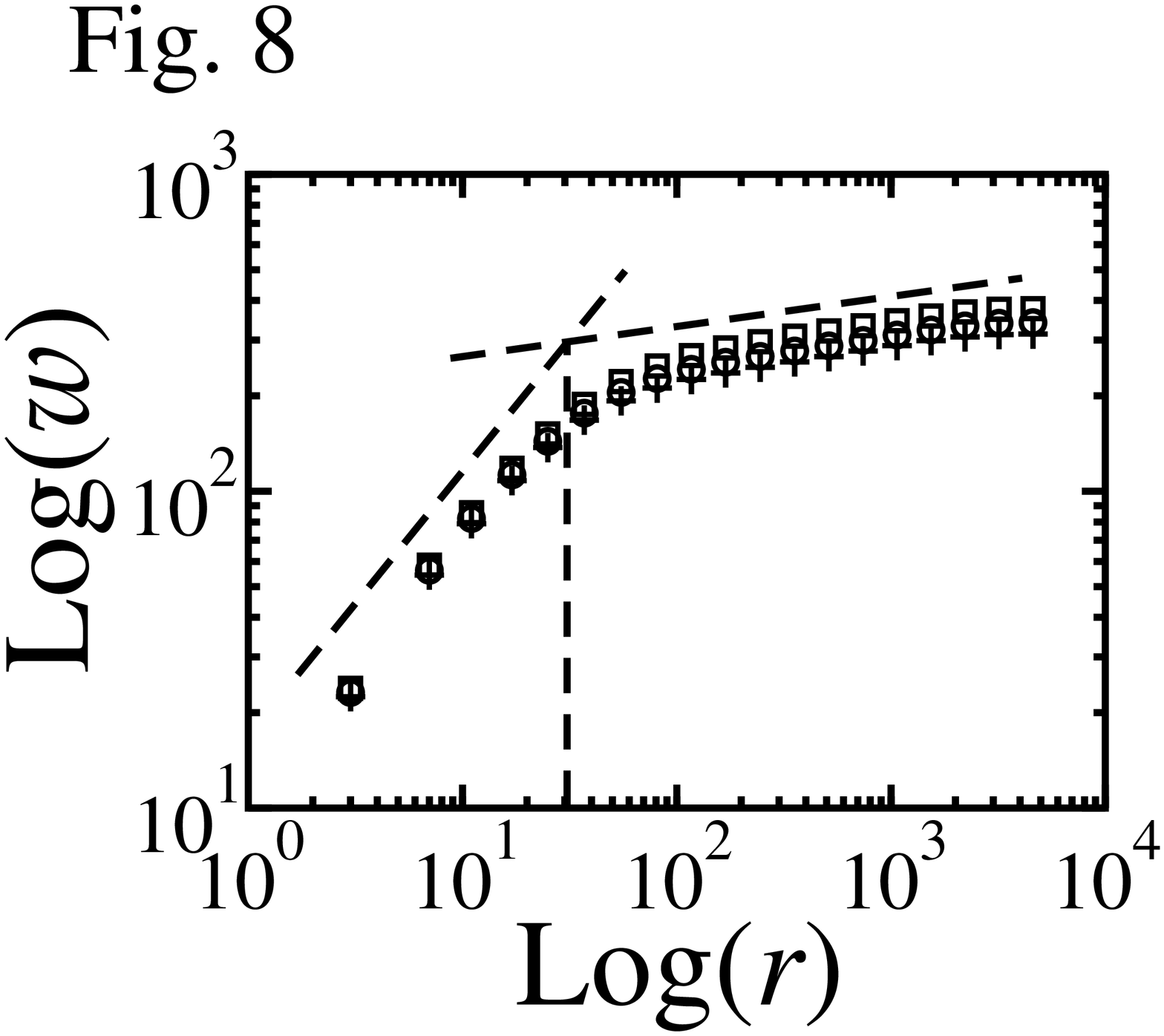}
\caption{Local roughness scaling for the grain deposition model with $\langle
l\rangle =32$ at time $t=2000$: Delta (crosses), Poisson (circles) and
Gaussian (squares) size distributions. The dashed lines are parallel to the
linear fits of two scaling regions, which give the estimate of $r_c$.}
\label{fig8}
\end{center}
\end{figure}

\begin{figure}[!h]
\begin{center}
\includegraphics[width=13cm]{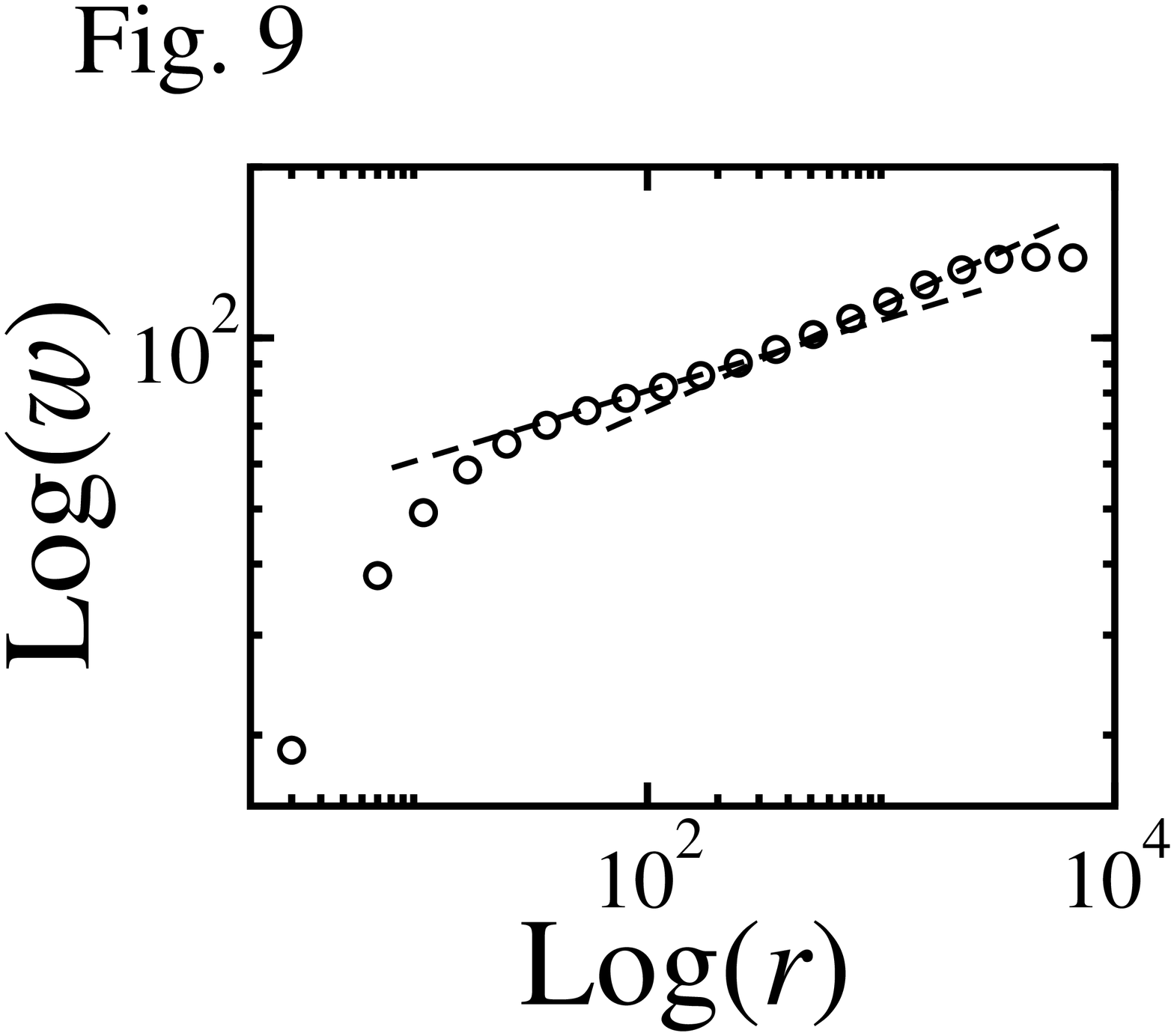}
\caption{Local roughness scaling for the grain deposition model with $\langle
l\rangle =8$ and Poisson size distribution for $t=6000$, where a change in
$\alpha_2$ is observed as the window size $r$ increases. Dashed lines are
linear fits of the data for two half-decades of $r$.}
\label{fig9}
\end{center}
\end{figure}

\begin{figure}[!h]
\begin{center}
\includegraphics[width=13cm]{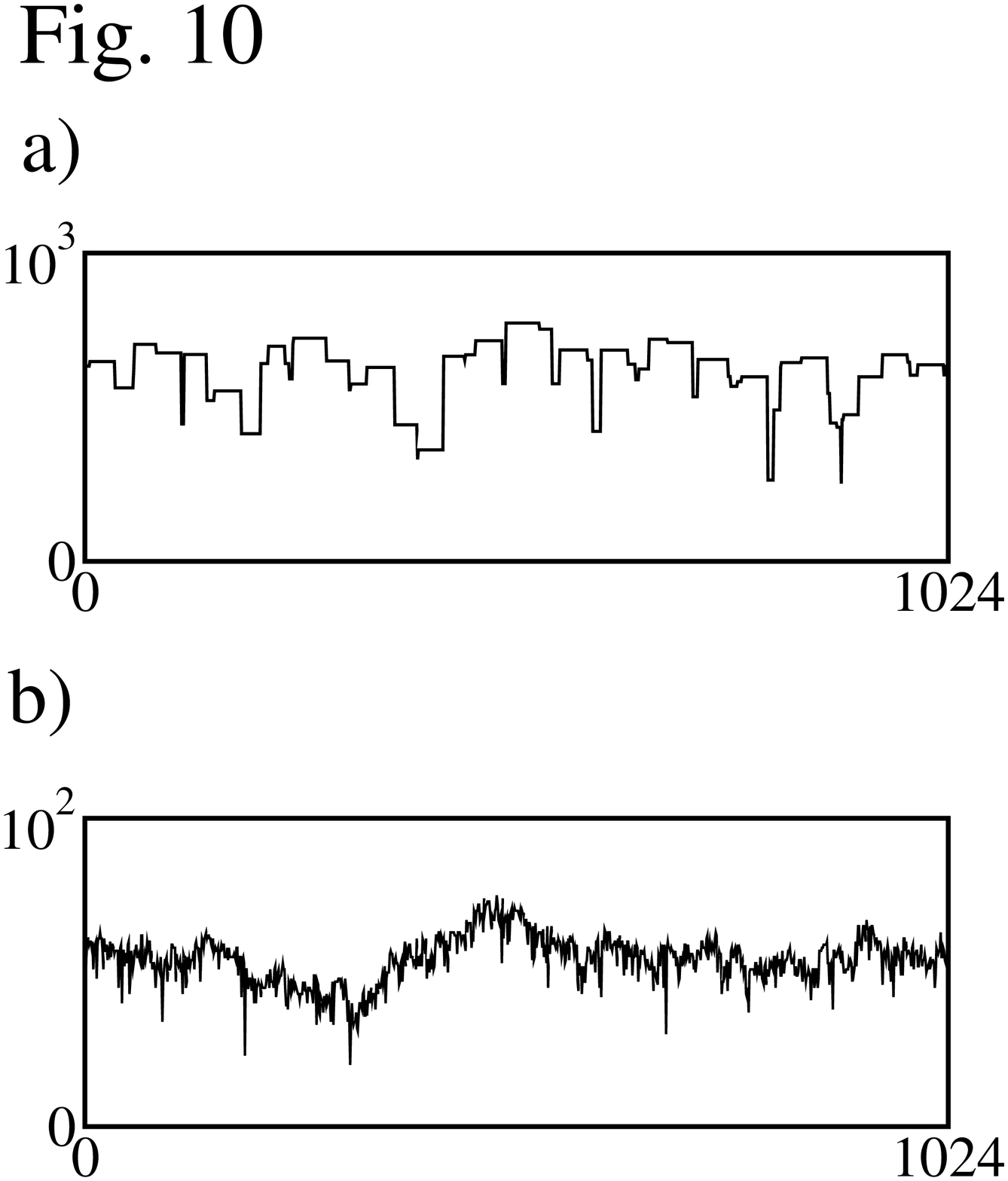}
\caption{(a) Cross-sectional view at of the surface of a deposit at $t=500$
grown with the grain deposition model with $\langle l\rangle =32$ and Gaussian
size distribution. (b) Cross-sectional view of the surface of a deposit at
$t=500$ grown with the ballistic deposition model.}
\label{fig10}
\end{center}
\end{figure}

\end{document}